\documentclass[usenatbib]{mnras}
\usepackage{graphicx}
\usepackage{graphics}
\usepackage{amsmath}
\usepackage{multirow}
\usepackage{color}
\usepackage{xcolor}
\usepackage{bm}		
\usepackage{float}

\title[47 Tucanae globular cluster]{Close encounters involving RAVE stars beyond the 47 Tucanae tidal radius.}

\author[J. G. Fern\'andez-Trincado et al.]{J. G. Fern\'andez-Trincado$^{1}$\thanks{Contact e-mail:
		\href{mailto:jfernandez@obs-besancon.fr}{jfernandez@obs-besancon.fr and/or jfernandezt87@gmail.com}},  
	A. C. Robin$^{1}$, C. Reyl\'e$^{1}$, K. Vieira$^{2}$, M. Palmer$^{3}$,\linebreak \newauthor  E. Moreno$^{4}$,  O. Valenzuela$^{4}$, B. Pichardo$^{4}$\\
	(Affiliations can be found after the references)}

\begin{document}
\label{firstpage}
\pagerange{\pageref{firstpage}--\pageref{lastpage}}
\maketitle

\begin{abstract}

The most accurate 6D phase-space information from the Radial Velocity Experiment (RAVE) was used to integrate the orbits of 105 stars around the galactic globular cluster 47 Tucanae, to look for close encounters between them in the past, with a minimum distance approach less than the cluster tidal radius. The stars are currently over the distance range 3.0 kpc $<$ d $<$ 5.5 kpc. Using the uncertainties in the current position and velocity vector for both, star and cluster, 105 pairs of star-cluster orbits were generated in a Monte Carlo numerical scheme, integrated over 2 Gyr and considering an axisymmetric and non-axisymmetric Milky-Way-like Galactic potential, respectively. In this scheme, we identified 20 potential cluster members that had close encounters with the globular cluster 47 Tucanae, all of which have a relative velocity distribution (V$_{rel}$) less than 200 km s$^{-1}$ at the minimum distance approach. Among these potential members, 9 had close encounters with the cluster with velocities less than the escape velocity of 47 Tucanae, therefore a scenario of tidal stripping seems likely. These stars have been classified with a 93\% confidence level, leading to the identification of extratidal cluster stars. For the other 11 stars, V$_{rel}$ exceeds the escape velocity of the cluster, therefore they were likely ejected or are unassociated interlopers.

\end{abstract}                
 


\section{Introduction}
\label{Section1:Introduction}

47 Tucanae is the second most massive/luminous ($\sim 1.1\times10^{6}$  M$_{\odot}$) Galactic globular cluster
\citep[][]{Lane20101}, one of the closest to the 
Milky May \citep[$\sim$ 4.02 kpc,][]{McLaughlin2006}, and 
``a test-bed for Galaxy formation models'', with a valuable reservoir of stars of different types 
useful to understand the formation process of multiple stellar populations in globular clusters. Recent studies on self-enrichment scenarios by 
massive AGBs, oriented to explain abundance patterns, support the idea that 47 Tucanae was possibly $\sim$7.5 times more massive than it
is now \citep[e.g.,][]{Ventura2014}. 47 Tucanae exhibits a split or spread main sequence of stars and multiple subgiant branches \citep{Anderson2009}, multiple red giants branches \citep[][]{Milone2012}, and a third population visible only in the subgiant branch \citep[][]{Ventura2014}. These multiple sequences have been detected from high-precision photometry.

During the last years, it has become clear that 47 Tucanae is not the only Galactic globular cluster exhibiting multiple stellar populations, but it shows unusual kinematic properties \citep{Lane2010a}. Recently detected high velocity dispersion profile at large radii, suggests an ongoing evaporation, and under this scenario escaping stars could form tidal tails \citep{Lane2012}. In this regard, 
extra-tidal stars associated with 47 Tucanae are yet to be uncovered \citep{Lane2012}, and their detection from colour-magnitude diagrams (CMD)
proves difficult due to strong pollution by the Small Magellanic Cloud (SMC) in the background \citep{Leon2000}.

In the present paper we characterize the orbits of a sample of stars in the RAVE survey around 47 Tucanae, with distances between 3.0 kpc to 5.5 kpc.
Under a Monte Carlo scheme, star-cluster orbits pairs are integrated in an axisymmetric and non-axisymmetric Milky-Way-like
Galactic potential, searching for close encounters in the past. The probability\footnote{The probability can be quantified in the same manner as in \citep{Fernandez-Trincado2015b}, i. e., $Prob =$ Number of orbits having close encounters with the cluster / $N_{total}$, where $N_{total} = 1\times10^{5}$ Monte Carlo simulations} of occurrence is evaluated for stars having close
encounters in the past with the cluster within such time. Two cases are proposed in our Monte Carlo scheme: (\emph{i}) Stars having encounters with relative velocity less 
than the cluster's escape velocity are considered to be likely tidal stripping, and (\emph{ii}) Stars with relative velocity greater than 
this escape velocity are considered to be likely ejected, which can be explained by some scenarios involving black 
holes, or binary system interactions \citep[e.g.,][]{Gvaramadze2009, Pichardo+2012, Lutzgendorf2012, Lind2015}.

This procedure is similar to the one by \citet{Fernandez-Trincado2015b, Fernandez-Trincado2015a} for globular cluster $\omega$ Centauri, and it is applied here for
the first time to 47 Tucanae. 
The dynamical history of clusters which are believed to have been more massive in the past, provides very useful information to understand the formation of the Galactic halo. To date, RAVE has gathered half-a-million medium resolution (R$\sim$7500) spectra of stars in the southern sky, 
and because RAVE input catalogue is magnitude limited (9 $<$ I $<$ 13), the stellar sample is essentially homogeneous
and free of kinematic biases. The advent of unprecedented accuracy 6D phase-space data from large survey such as Gaia,
fed into Monte Carlo simulations like the one presented in this paper, will be particulary valuable to help us characterize 
the kinematics of stars in and out the tidal radius of Galactic globular clusters. 

In this paper we test the viability to using the integration of orbits in a Monte Carlo numerical scheme, looking at the relative velocity of enconters ocurred within the cluster tidal radius, which has been extensively tested with the results of \textit{N-}body simulations.  The paper is structured as follows: In Section \ref{selection}, we briefly describe the accurate RAVE data set available in $\sim120$ sq. deg. centred on 47 Tucanae; in Section \ref{simulation},  we describe the Monte Carlo numerical scheme used to constrain potential cluster members; results and discussion are presented in Section \ref{results} and in Section \ref{conclusion}, we summarize the main conclusions of this paper.


\begin{figure}
	\begin{center}
		\includegraphics[width=95mm,height=85mm]{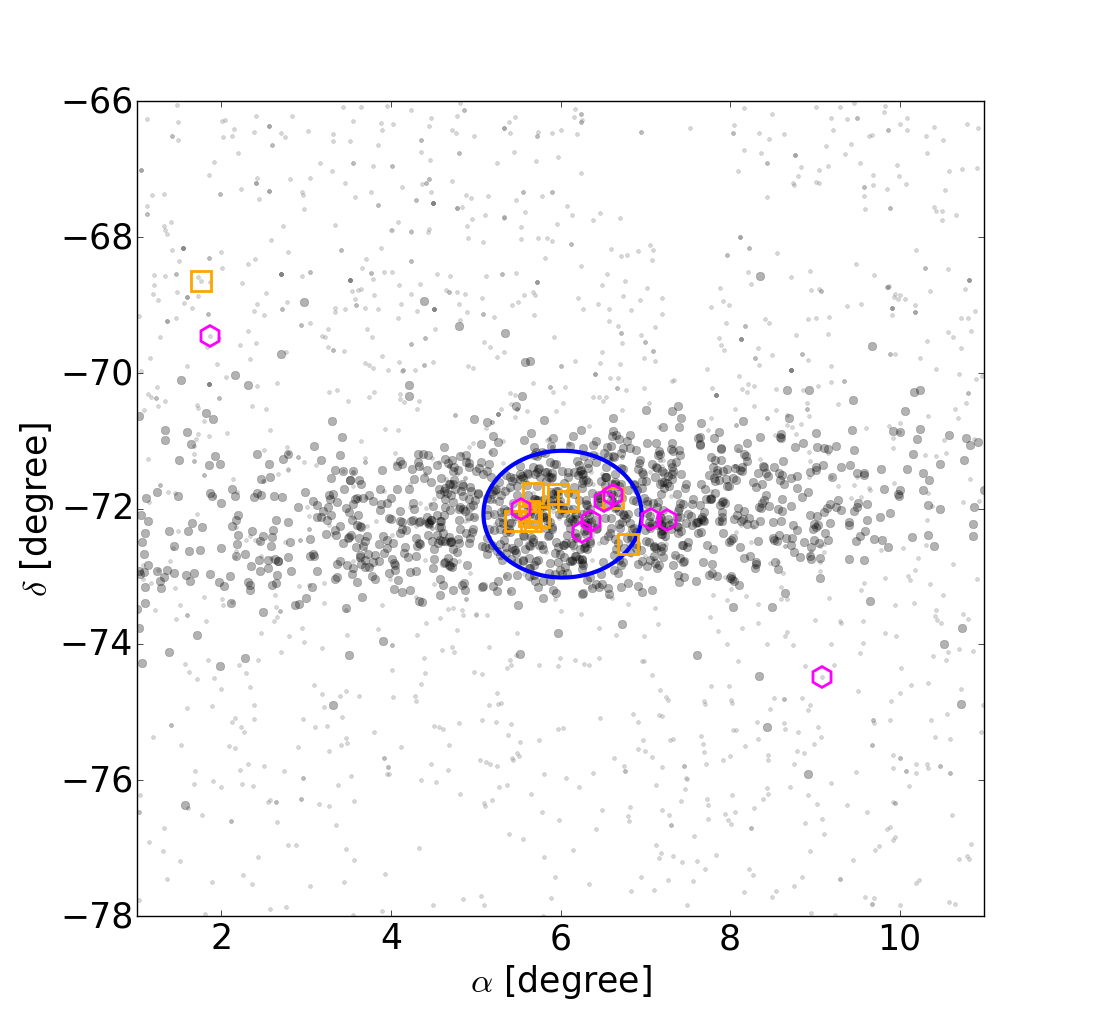}
	\end{center}
	\caption{Spatial distribution for 20 potential extra tidal stars associated with 47 Tucanae. Magenta open symbols mark the positions of stars candidates with $V_{rel}< 68.8$  km s$^{-1}$, orange open symbols marks the positions of stars candidates with $V_{rel}> 68.8$  km s$^{-1}$. All these stars are currently well outside 47 Tucanae tidal radius (r$_{t} > 56$ pc), whose projection on the sky is shown with the large blue circle shows. The black dots are stars in the RAVE survey in the same direction of the cluster. The grey filled dots show the present-day expected angular distribution of debris stars as predicted by \textit{N}-body particles around 120 deg$^{2}$ area centered on 47 Tucanae. (A color version of this figure is available in the online journal).} 
	\label{Figure1}
\end{figure}

\begin{figure*}
	\begin{center}
		\includegraphics[width=180mm,height=60mm]{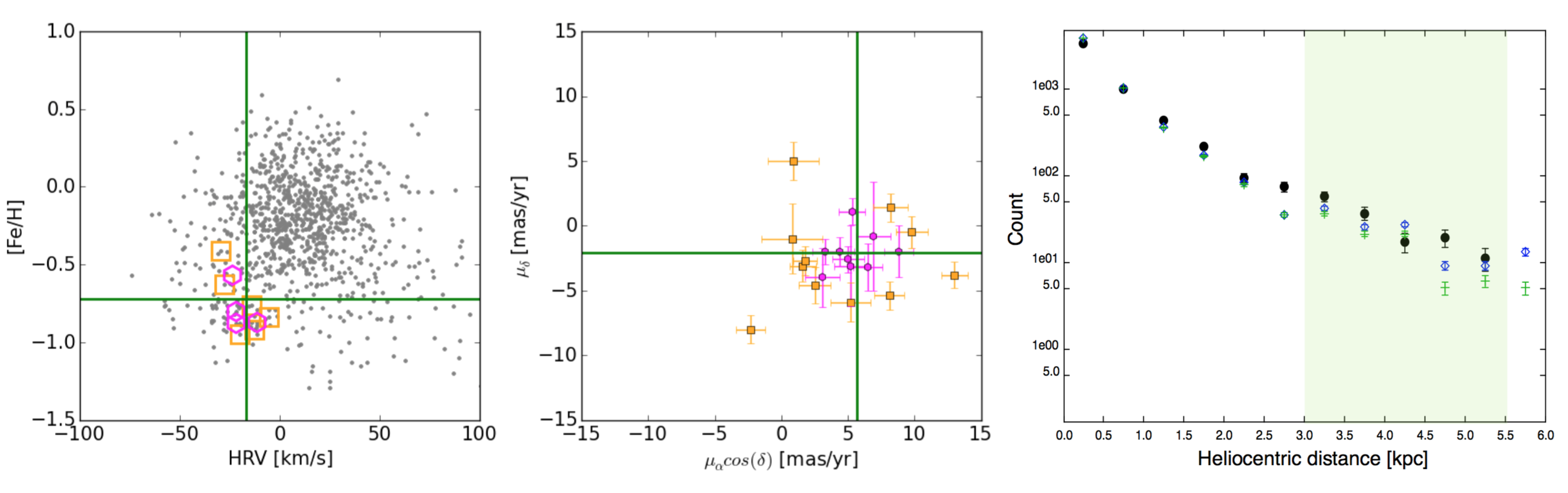}
	\end{center}
	\caption{Left panel: Radial velocity-metallicity distribution for stars candidates similar to the nominal radial velocity and metallicity ([Fe/H]) of 47 Tucanae (horizontal and vertical lines). Grey dots mark RAVE field stars in the same direction of the cluster (see Figure \ref{Figure1}). Central panel: Proper motions distribution for stars with relative velocity less (magenta symbols) and greater (orange symbols) than the escape velocity of the cluster, as in Figure \ref{Figure1}. The nominal proper motions of the cluster is marked with the green horizontal ($\mu_{\delta}$) and vertical ($\mu_{\alpha}cos(\delta)$) lines. Right panel: Expected cumulative distribution of heliocentric distances for foreground stars from the Galaxy, obtained from mock samples from the Besan\c{c}on Galaxy model (blue diamonds). The green crosses shows the contribution of the Galactic disk. The black full circles shows the cumulative distribution of heliocentric distances from the RAVE sample in the same direction. The green region indicates the established limits in this work to identify extra tidal stars. (A color version of this figure is available in the online journal).} 
	\label{Figure2}
\end{figure*}

\section{Sample selection}
\label{selection}

We selected a subsample of candidates stars for a 120 deg$^2$ area centered on 47 Tucanae (see Figure \ref{Figure1})
from the RAVE DR4 public data \citep{Kordopatis+2013}, 
with very accurate radial velocity (signal-to-noise ratio better than 20, allows us to ensure radial velocity errors  $<$ 2 km s$^{-1}$; \citep[e.g.,][]{Hawkins+2015}. We assumed the distance estimates based on the Bayesian approach of \citet{Binney2014}, with errors less than 20\%. The RAVE DR4 catalog is complemented with several proper motions in the literature, however, 
we have decided to use the UCAC4 proper motions \citep{Zacharias+2013} because those are available for all our sample, and the proper motions are less affected by potential systematic uncertainties than in other catalogues \citep{Vickers2016}.

The systematic radial velocity of 47 Tucanae is not enough to separate likely cluster members 
from Galactic foreground stars \citep[e.g.,][]{Meylan1991}. 
To minimize Galactic contamination, we select stars with heliocentric distances between 3.0 kpc to 5.5 kpc. 

Theses limits in distance were adopted based on the distance spread of the expected tidal debris stars as predicted by an \textit{N}-body simulation of 47 Tucanae orbiting the Milky Way \citep{Lane2012}. Figure \ref{Figure1} shows the present-day angular positions of \textit{N}-body particles.

We explore the expected Galactic contamination based on the updated version of the Besan\c{c}on Galaxy model \citep{Robin2014}. From the right panel of Figure \ref{Figure2}, one can estimate that the Galactic disk and halo contributes $\sim$ 1.7\% and $\sim$ 1.0\% respectively, of all stellar population for distances greater than 3.0 kpc of the Sun, and 0.1\% and 0.6\% for distances greater than 5.5 kpc. An excess of stars between 3 kpc and 5 kpc from the RAVE DRA4 as compared to the Besan\c{c}on Model using the same selection function in both datasets, is naturally caused by the presence of the cluster in the observed field.

Our final sample has 105 RAVE stars with 6D phase-space information.

\section{Milky-Way-like Galactic model and orbits}
\label{simulation}

In the orbits computations, we used semi-analytic Milky Way gravitational potential models, including axisymmetric and non-axisymmetric ones, where dynamical friction is ignored. For the non-axisymmetric Galactic model, we first scaled the \citet{Allen+1991} bulge-disk-halo axisymmetric model following a Monte Carlo scheme, taking the Local Standard of Rest (LSR) parameters (R$_{0}$,$\Theta_{0}$)= (8.3$\pm$0.23 kpc, 239$\pm$7 km s$^{-1}$) given by \citet{Brunthaler+2011}, and proceed to transform all the mass of the scaled bulge into a prolate or boxy bar following the procedure given by \citet{Pichardo+2004}. The bar component approximates the three dimensional mass density distribution of the Freudenreich (1998) Galactic bar model, based on COBE/DIRBE observations of the Galactic center. In addition, our model includes the three dimensional model for a two-armed spiral component, called PERLAS, given by \citep{Pichardo2003}; these arms are built with a small fraction of the mass of the scaled disk. The relevant parameters employed in the bar and the spiral arms are the following: the pitch angle of the spiral arms is \textit{i} $=$ 15.5$\pm$3.5, and the initial angle between the bar’s major axis and the Sun-Galactic center line is $ = 20$ deg; the ratio of the mass in the spiral arms to the mass of the disk component is $M_{arms}/M_{disk} = 0.04\pm0.01$; the angular velocities of the bar and spiral arms are respectively $\Omega_{bar}=55 \pm 5$ km s$^{-1}$ kpc$^{-1}$, $\Omega_{arms}=25\pm5$ km s$^{-1}$ \citep{Moreno+2014}.  

It is important to note that small variations in the form of the potential do not lead to significant changes in the properties of the computed orbits (see Figure \ref{Figure3}). Table \ref{table2} show the comparison in both Galactic potentials (axisymmetric and non-axisymmetric structures) of the relative velocity between the stars and 47 Tucanae.

To study the kinematics of our sample in the Milky Way, we compute the orbits considering a series of Monte Carlo sampling simulations. We employ the six-dimensional phase space coordinates of these stars and those associated with the Galactic model, and consider their corresponding uncertainties as 1$\sigma$ variations, plus the uncertainties of the Galactic parameters in a Gaussian Monte Carlo sampling in the same manner as \citet{Moreno+2014}. Each run has its corresponding sampled Galactic model and star's initial orbital conditions. The adopted Solar motion with its uncertainties is ($U, V, W$)$_{\odot} = (-11.1\pm1.2, 12.24\pm2.1, 7.25\pm0.6)$ km s$^{-1}$ \citep{Schonrich2010, Brunthaler+2011}

Computations considered the
effect of the quoted uncertainties in the position and velocity of the
LSR on the Galactic potential. These uncertainties are taken as 1$\sigma$ variations, and a Gaussian Monte Carlo sampling generates
the parameters to compute the present-day positions and velocities of the cluster and star.
Uncertainties in the solar motion are also taken into account.
Our Monte Carlo numerical scheme used all these values \citep[e.g.,][]{Pichardo+2012, Moreno+2014}.

For 47 Tucanae itself, we calculated the orbit based on its current distance, proper motions, radial velocity, and positions
as listed in Table \ref{table1}. For the selected sample, the 6D phase-space information is used in the Monte Carlo numerical scheme, 
taking into account the observational uncertainties of $R_0$, $\Theta_0$, and RAVE data. 
We ran $10^5$ star-cluster pair of orbits, looking for close encounters 2 Gyr backwards in time. 
This would be the time needed to build up tidal tails from the cluster \citep[e.g.,][]{Lane2012}, 
detectable today as RAVE stars associated with 47 Tucanae beyond the cluster's tidal radius.

The minimum approach distance $d_{min}$, is the straightforward indicator of close encounters 
between RAVE stars and 47 Tucanae along their orbits, e.g., if at the time of the encounter 
the star is within the tidal radius of 47 Tucanae ($d_{min} <  56 $ pc). The adopted value ($d_{min}$) is in agreement with the data from \citet{Harris1996}, 
where the surface density drops to zero at 56 pc from the centre of the cluster for an adopted King profile. 

Then, encounters with the adopted $d_{min}$ and relative velocity smaller that the escape velocity, 
$V_{0} = 68.8$ kms$^{-1}$, are considered as ``tidally stripped" from the cluster, 
while encounters with relative velocity larger that the escape velocity and relative velocity $V_{rel}<200$ kms$^{-1}$, 
are consistent with stars being ''ejected`` from the cluster by means of three-body encounters or black hole interactions 
\citep[e.g.,][ see references therein]{Leonard1991, Pichardo+2012, Lind2015}.  Figure \ref{Figure3} shows the relative velocity ($V_{rel}$)
distribution functions for 20 potentially members of 47 Tucanae.

\begin{table*}
	\setlength{\tabcolsep}{6mm}  
	\caption{Positions and velocities of 47 Tucanae.}
	\begin{tabular}{llc}
		\hline
		\hline
		Parameter            & Value   & Reference \\
		\hline
		\hline
		Distance from the Sun              &    $4.02 \pm 0.35$ kpc & \citet{McLaughlin2006}\\
		$\alpha (J2000)$               &   $00:24:05.67$ & \citet{McLaughlin2006}\\
		$\delta (J2000)$                &   $-72:04:52.62$ & \citet{McLaughlin2006}\\
		$\mu_{\alpha} cos(\delta)$ &  ($5.64 \pm 0.20$) mas yr${-1}$& \citet{Anderson2004}\\
		$\mu_{\delta}$         & ($-2.05 \pm 0.20$) mas yr${-1}$& \citet{Anderson2004}\\
		$V_r$                    & ($-16.85 \pm 0.16$) km s${-1}$ & \citet{Lane2010a}\\
		$r_t$ (tidal radius)                     & 56 pc & \citet{Harris1996} \\
		$V_{0}$ (escape velocity)  & 68.8 km s$^{-1}$ & \citet{Gnedin2002} \\
		\hline
		\hline
	\end{tabular}  \label{table1}
\end{table*}

In particular, we analyzed possible encounters in the recent 300 Myr; 
in this time interval, 47 Tucanae has about one perigalactic passage, since
the orbital period reported for the cluster spans between 190$\pm$4 to 193$\pm$4 Myr \citep[e.g.,][]{Dinescu1999}.

\subsection{Energy (\emph{E}) and angular momentum ($L{z}$) from the Monte Carlo numerical scheme}

An important point must be noted about the Monte Carlo scheme used in this investigation:
A star may have been ejected from 47 Tucanae, and its resulting energy ($E$) and angular momentum ($Lz$)
from the stripping/ejection process may be completely different from the progenitor cluster. 
To illustrate this point, we explain the following example: 
the energy and angular momentum of 47 Tucanae are simulated taking into account a Monte Carlo gaussian sampling 
of the cluster's radial velocity, distance, proper motions, etc., within their quoted errors.
Then, a set of 10$^{5}$ Monte Carlo $L_z$ versus $E$ points are generated, 
producing the distribution of grey dots observed in Figure \ref{ELz}, 
where the blue symbol refers to the average value 
$\langle E \rangle = -1486$ km$^{2}$s$^{-2}$ and $\langle L_z \rangle = 1389$ kms$^{-1}$kpc. 
In the same figure, four ellipses show the 1-, 2- ,3- and 4-$\sigma$ regions of the distribution. 
We first consider stars being ejected from the cluster with a relative velocity of
$20$ kms$^{-1}$, $40$ kms$^{-1}$, and $60$ kms$^{-1}$ towards many different directions
defined by a fine grid of points on the celestial sphere centered at the cluster.
We compute for each direction the values $E$ and $L_z$ of the ejected star (magenta dots in Figure \ref{ELz}). 
It is clear from Figure \ref{ELz} that many of these points 
may lie far from the 4$\sigma$ curve, and then can be very different from the values ($L_z$, $E$) of the cluster. 
Based on the above discussion, the value of $E$ and $L_z$ cannot be used as a sufficient conditions to identify
stars escaping from 47 Tucanae in our Monte Carlo simulations.\\

\subsection{\textit{N-}body simulations as testbed in the integration of orbits using a Monte Carlo numerical scheme}

To supplement our results, we also investigated the orbits for a suite of \textit{N-}body simulations of satellites undergoing tidal disruption along 47 Tucanae-like orbits (present-day) in the Milky Way potential \citep[e.g.,][]{Lane2012}.  We have selected randomly 500 particles (see Figure \ref{Figure1}) and randomly assigned observational errors, to account for observational uncertainties that prohibit us from accurately knowing the true position, space motion,	and therefore orbits of both 47 Tucanae and the 500 particles. For each particle we integrate $1\times{}10^5$ orbits looking for close enconters in the same way as already mentioned above. For the Galactic gravitational potential we employed our proposed non-axisymmetric mass model. The resulting 
relative velocity ($V_{rel}$) distribution function of \textit{N-}body particles are also shown in Figure \ref{Figure7} and Figure \ref{Figure8}. 

We note that as we start with the current structure of the cluster, which is then tidally stripped along its orbit, the final cluster system in our model has slightly different properties than 47 Tucanae, but our Monte Carlo approach for orbits integration is still effective to identify close enconters between the \textit{N-}body particles and 47 Tucanae-like.	

We find that 93\% of the \textit{N-}body particles encounters occurred within the tidal radius of 47 Tucanae and do not exceed their escape velocity. These results allow us clearly quantify the confidence level of our technique of integration using a Monte Carlo approach to find probable extra tidal stars candidates nearly 47 Tucanae. The initial conditions (\textit{N-}body particles) follow the typical observed uncertainties of stars observed in RAVE survey in the same field, i.e., the errors in proper motions are assumed to be $<5$ mas/yr, in radial velocity $<10$ km s$^{-1}$, and the uncertainties in distances are on the order of 20$\%$. Under this assumption we demonstrate with a $93\%$ of confidence level that \textit{N-}body particles associated to disrupting particles from 47 Tucanae, can be identified as potentially members.  

To facilitate the reproducibility and reuse of our results, we have made the Monte Carlo simulations available in a public repository at \url{https://github.com/Fernandez-Trincado/47Tucanae-NbodySimulations}.

\begin{figure*}
	\begin{center}
		\includegraphics[width=190mm,height=210mm]{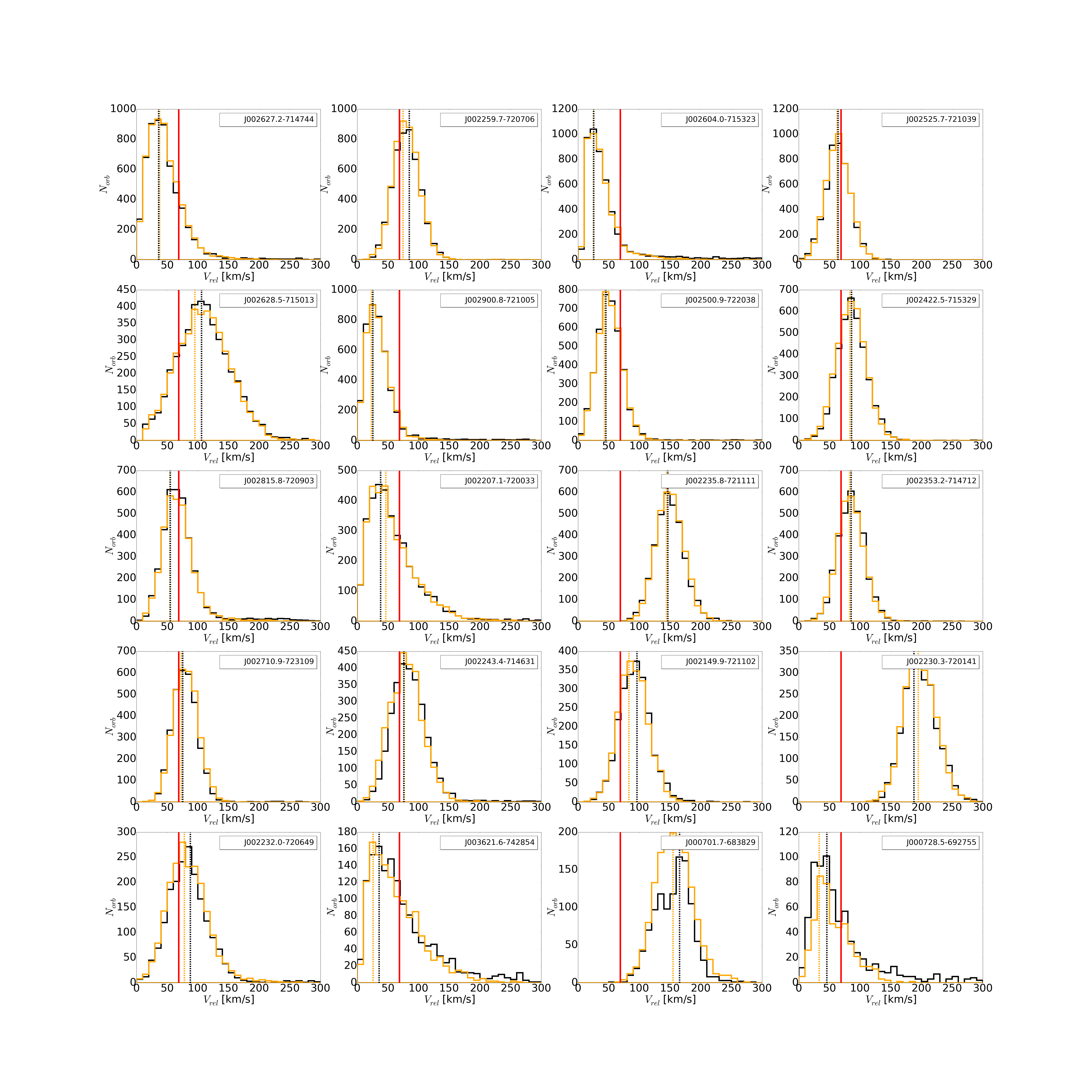}
	\end{center}
	\caption{Relative velocity ($V_{rel}$) distribution funtions for 20 stars having close encounters with 47 Tucanae (see Section \ref{simulation}) in an axisymmetric Galactic potential (black line), and non-axisymmetric (orange line) Galactic potential of the Milky Way.  The vertical red line marks the escape velocity from 47 Tucanae, $V_{0}=68.8$ km/s$^{-1}$, and the vertical dashed line refers to the more frequent relative velocity listed in Table \ref{table2}. The distributions are accumulated from $10^5$ Monte Carlo realizations of the clusters and star orbits. (A color version of this figure is available in the online journal).} 
	\label{Figure3}
\end{figure*}

\section{Results and discussion}
\label{results}

Figure \ref{Figure3} shows the relative velocity distribution funtions ($V_{rel}$) during close encounters (star - cluster) for 20 potential members of 47 Tucanae. We found that 45\% (9 stars) of those stars may have escaped from the cluster with a velocity less than the escape velocity, satisfying a possible scenario of tidally stripping, while 55\% of the stars listed in Table \ref{table2} seems likely ejected from the cluster with velocities greater than 
the escape velocity. Also, Figure \ref{Figure3} shows that moderate changes in the form of the Galactic potential and the 47 Tucanae gravitational force\footnote{A static potential assuming a King profile \citep{King1962} has been considered for the cluster itself.} itself result in non-significant changes in our predicted results (see Table \ref{table2}).

\begin{figure}
	\begin{center}
		\includegraphics[width= 90mm,height=135mm]{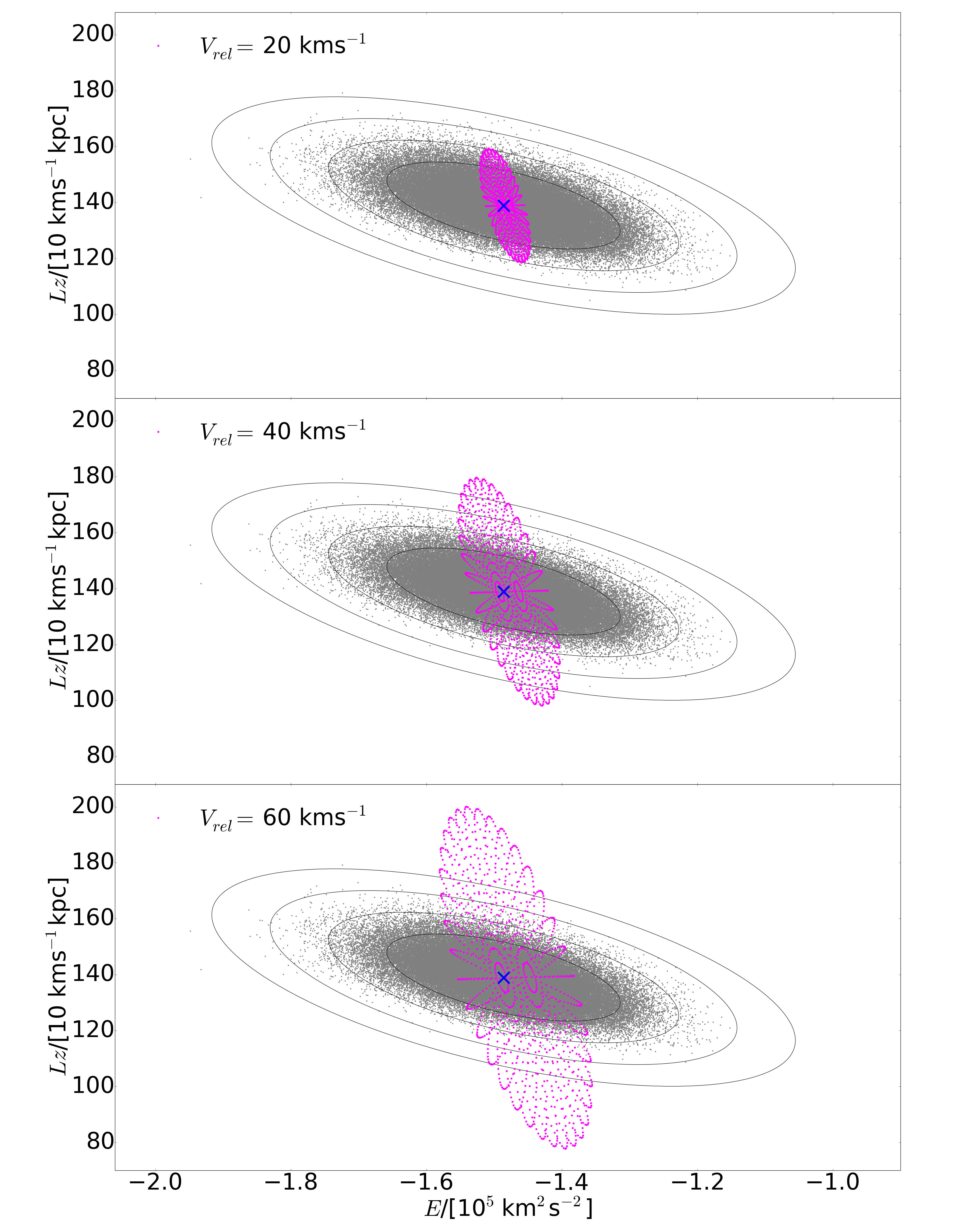}
	\end{center}
	\caption{Orbital angular momentum ($L_z$) and energy ($E$) generated from 10$^{5}$ Monte 
		Carlo simulations of 47 Tucanae (grey dots), taking into account Gaussian uncertainties. Magenta dots are simulated stars escaping from the cluster with relative velocities of 20 kms$^{-1}$ (top panel), 40 kms$^{-1}$ (middle panel) and 
		60 kms$^{-1}$ (bottom panel). The blue symbol refers to the average values $L_z$ and $E$ from the distribution (grey dots) for the concentric
		ellipses (black lines) with 1$\sigma$, 2$\sigma$, 3$\sigma$, and 4$\sigma$. (A color version of this figure is available in the online journal).} 
	\label{ELz}
\end{figure}

The RAVEID, number of Monte Carlo orbits having close encounters, the most frequent relative velocity, and the most frequent encounter time ($t_{enc}$), are listed in Table \ref{table2}.

We identified 9 stars in the direction of the 47 Tucanae globular cluster with probabilities greater than 0.5\% having been ejected from the cluster, with relative velocities less than escape velocity, $V_{rel}<V_{0}$, within the past 2 Gyr. Our identification is based on the integration of orbits under a Monte Carlo approach, which allowed us to classify potential members 
with a 93\% of confidence level. Figure \ref{Figure1} shows that these stars appear to have the same [Fe/H] metallicities, radial velocity, and proper motions than 47 Tucanae. We also note that four stars (J000728.5$-$692755, J002815.8$-$720903, J002900.8$-$721005, J003621.6$-$742854)  outside of the tidal radius have similar proper motions to the cluster, and are potential candidates for membership.

We find 11 stars in our sample that exceeds the escape velocity of the cluster, $V_{rel}>V_{0}$, during the close encounters with 47 Tucanae. This velocity can be explained by some peculiar scenarios containing black holes and/or binary system. However, we noted that these velocities were not observed in our \textit{N}-body particles experiment, which may be because the \textit{N}-body simulations do not incorporate binary systems. We evaluated every observable and found that relatively large proper motions, shown as orange symbols in Figure \ref{Figure2}, can produce large observed relative velocity, therefore significantly changing one of the components of the orbital velocity vector. Our results cannot rule out the possibility that these stars are associated with the 47 Tucanae progenitor system or if they are unassociated interlopers, whose orbits coincide with the cluster's one, Only the determination of accurate chemical abundances could help to constrain true cluster members in this scenario. We will investigate in more detail the physical process that imprint out these high velocities, on a large sample of other Galactic globular cluster with extra tidal stars candidates. With the very accurate radial velocity, distance and proper motions from the upcoming six-dimensional phase- space data set that will be produced by the Gaia space mission, we will be able to do a more complete characterization of the velocity vector.

\begin{table}
	\setlength{\tabcolsep}{6.0mm}  
	\begin{tiny}
		\caption{Ejection parameters for 20 more likely RAVE stars having close encounters with 47 Tucanae, obtained with the axisymmetric (first line) and non-axisymmetric (second line) potentials. For each star the ID from RAVE catalogue and the number of orbits ($N_{orb}$ ) having close encounters with the cluster are given in column 1, and 2. The relative velocity ($V_{rel}$) and the encounter time are given in column 3, and 4.}
		\label{table2}
		\begin{tabular}{lccc}
			\hline
			\hline
			RAVEID &   $N_{orb}$  &  $V_{rel}$     &  $\tau_{enc}$\\  
			&                     &  km s$^{-1}$ & Myr \\
			\hline
			\hline
			$V_{rel}<V_{0}$    &        &         &        \\
			\hline
			J002627.2$-$714744 &  5748  &   36.15 &  0.499 \\
			&  5835  &   37.61 &  0.146 \\
			J002604.0$-$715323 &  4815  &   25.52 &  0.498 \\
			&  4588  &   25.01 &  0.146 \\
			J002525.7$-$721039 &  4700  &   63.74 &  0.498 \\
			&  4797  &   62.29 &  0.141 \\
			J002900.8$-$721005 &  4170  &   25.28 &  0.499 \\
			&  4007  &   22.80 &  0.146 \\
			J002500.9$-$722038 &  3968  &   45.33 &  0.498 \\
			&  3898  &   42.97 &  0.146 \\
			J002815.8$-$720903 &  3634  &   54.93 &  0.498 \\
			&  3440  &   54.22 &  0.146 \\
			J002207.1$-$720033 &  3484  &   37.90 &  0.498 \\
			&  3449  &   46.50 &  0.146 \\
			J003621.6$-$742854 &  1435  &   35.57 &  2.499 \\
			&  1330  &   25.66 &  1.146 \\
			J000728.5$-$692755 &  717   &   45.52 &  2.499 \\
			&  479   &   33.09 &  2.146 \\
			\hline
			$V_{rel}>V_{0}$    &        &         &        \\
			\hline
			J002259.7$-$720706 &  4842  &   84.63 &  0.482 \\
			&  4979  &   74.41 &  0.143 \\
			J002628.5$-$715013 &  4313  &   106.0 &  0.490 \\
			&  4262  &   95.07 &  0.146 \\
			J002422.5$-$715329 &  3776  &   85.90 &  0.498 \\
			&  3893  &   83.21 &  0.145 \\
			J002235.8$-$721111 &  3424  &  146.29 &  0.497 \\
			&  3491  &  144.03 &  0.138 \\
			J002353.2$-$714712 &  3196  &   85.26 &  0.496 \\
			&  3140  &   82.78 &  0.146 \\
			J002710.9$-$723109 &  3180  &   75.15 &  0.494 \\
			&  3313  &   72.97 &  0.146 \\
			J002243.4$-$714631 &  2839  &   75.74 &  0.499 \\
			&  2911  &   76.45 &  0.146 \\
			J002149.9$-$721102 &  2299  &   95.99 &  0.497 \\
			&  2305  &   82.84 &  0.139 \\
			J002230.3$-$720141 &  2237  &  187.68 &  0.498 \\
			&  2273  &  194.81 &  0.001 \\
			J002232.0$-$720649 &  1944  &   87.79 &  0.499 \\
			&  2097  &   77.95 &  0.146 \\
			J000701.7$-$683829 &  1123  &  165.39 &  1.498 \\
			&  1552  &  154.66 &  1.001 \\
			\hline
			\hline
		\end{tabular} 
	\end{tiny}
\end{table}

For the remainder of the studied sample (85 RAVE stars) 
we have not found a significant number of close encounters with 47 Tucanae ($<$ 0.01\% probability or $<$ 10 orbits).

Assuming that our sample (stars listed in Table \ref{table2}) are true extra tidal stars associated to 47 Tucanae, it is interesting to compare their metallicity distribution ([Fe/H]) with that of cluster´s stars (rt $< $56 pc) and field stars. The metallicity distribution observed in the left panel of Figure \ref{Figure1}, of our candidates listed in Table 2, is comparable to the nominal [Fe/H]$\sim$ −0.72 for this cluster (Harris 1996). Figure \ref{Figure1} indicates that these stars candidates, as a group, have similar abundances [Fe/H] to 47 Tucanae, and fall approximately 0.5 dex from the mean metallicity of field stars in the same part of the sky.

\begin{figure*}
	\begin{center}
		\includegraphics[width=210mm,height=200mm]{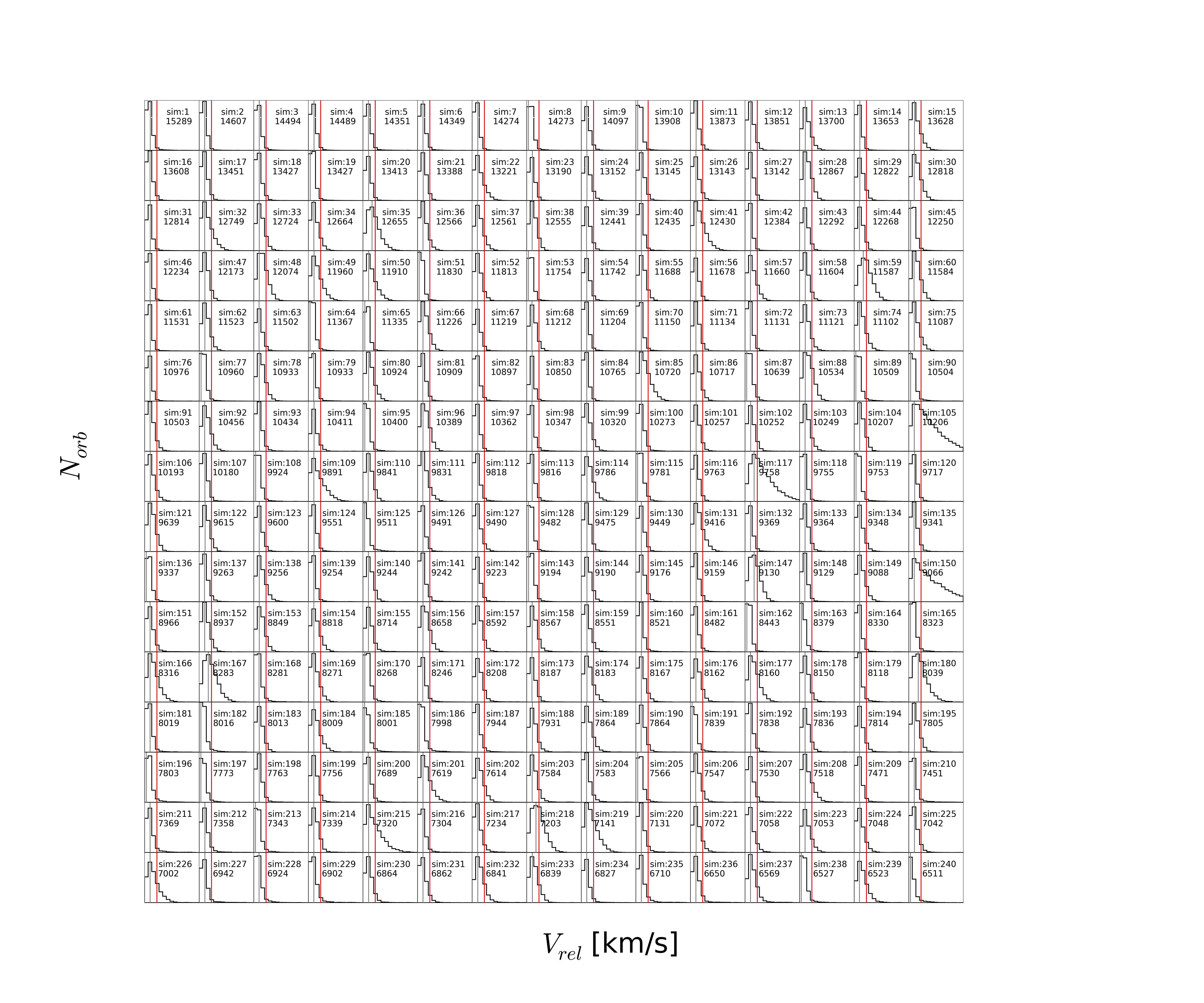}
	\end{center}
	\caption{Relative velocity ($V_{rel}$) distribution functions for close encounters between particles from \textit{N-}body simulations of satellites undergoing tidal disruption along 47 Tucanae \citep{Lane2012}. The vertical line indicates the escape velocity from the cluster (red line) and the more probable relative velocity of the particle (grey line). The scale is the same in these frame, with the vertical axis normalized to unity, and the horizontal axis spans relative velocities from 0 km s$^{-1}$ to 300 km s$^{-1}$.  (A color version of this figure is available in the online journal).} 
	\label{Figure7}
\end{figure*}

\begin{figure*}
	\begin{center}
		\includegraphics[width=210mm,height=200mm]{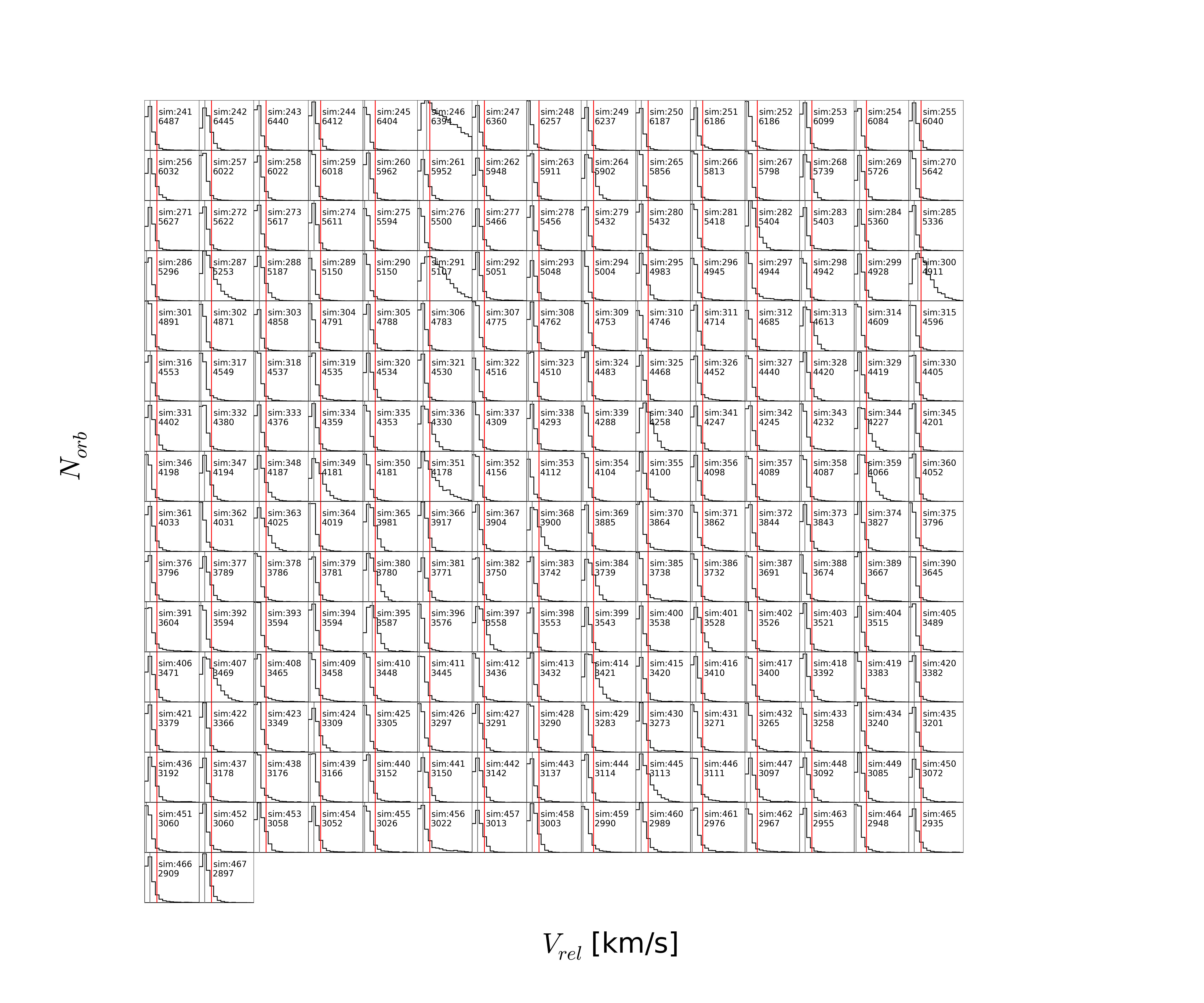}
	\end{center}
	\caption{Same as Figure \ref{Figure7}, but for simulations between 241 and 467. (A color version of this figure is available in the online journal).} 
	\label{Figure8}
\end{figure*}

\section{Conclusion}
\label{conclusion}

In this study, we have used for the first time the 6D phase-space data around Galactic globular cluster 47 Tucanae 
to compute pairs of star-cluster orbits in a Milky Way axisymmetric and non-axisymmetric Galactic potential following a kinematic Monte Carlo approach, 
looking for close encounters between the globular cluster and RAVE stars. Under this scheme, we constrain potential cluster former members based 
in their kinematic information and uncertainties.
We identified 20 potential extra tidal stars,
with kinematics and chemical pattern ([Fe/H]) similar to 47 Tucanae, and most likely ejected from the cluster 5 Myr ago (encounter times, $\tau$, are listed in Table \ref{table2}). We find with a $93\%$ of confidence level that 9 of those stars could have been tidally stripped from 47 Tucanae, based on the frequency of close enconters and the low relative velocity observed during each approach. In the near future, high resolution spectroscopy will be used to confirm such association.

Throughout this work we demonstrate that a Monte Carlo approach to integrate pairs of orbits (star - cluster's) can be used with a high level of confidence to identify extra-tidal stars candidates formerly associated with the cluster, assuming a scenario of tidal stripping, where the relative velocity is less than the escape velocity of the cluster. This hypothesis has been extensively used in stellar debris studies of Galactic globular clusters systems \citep{Lind2015, Fernandez-Trincado2015a, Fernandez-Trincado2015b, Fernandez-Trincado2016}. Here we extend its use to identify former members of a cluster that are now fully accreted into the Galactic halo.

We have found that the imprints of a rotating bar in a non-axisymmetric Galactic model does not affect the results of our simulations, and its influence can be neglected, in the problem being investigated in this paper.

\section*{Acknowledgements}

We thank the anonymous referee  for a detailed and constructive report which has improved the clarity of this work.

\textit{N-}body particles was kindly shared with us by Andrea Kunder.

J.G.F-T is currently supported by Centre National d'Etudes Spatiales (CNES) through PhD grant 0101973 
and the R\'egion de Franche-Comt\'e and by the French Programme National de Cosmologie et Galaxies (PNCG). 

E.M, and B.P acknowledge support from UNAM/PAPIIT grant IN105916. D.A.G.H was funded by the 
Ram\'on y Cajal fellowship number RYC-2013-14182.

Funding for RAVE has been provided by: 
the Australian Astronomical Observatory,
the Leibniz-Institut fuer Astrophysik Potsdam (AIP),
the Australian National University,  
the Australian Research Council, 
the French National Research Agency,
the German Research Foundation (SPP 1177 and SFB 881), 
the European Research Council (ERC-StG 240271 Galactica),
the Istituto Nazionale di Astrofisica at Padova, 
the Johns Hopkins University, 
the National Science Foundation of the USA (AST-0908326), 
the W. M. Keck foundation,
the Macquarie University,
the Netherlands Research School for Astronomy,
the Natural Sciences and Engineering Research Council of Canada,
the Slovenian Research Agency,
the Swiss National Science Foundation, 
the Science \& Technology Facilities Council of the UK,
Opticon, Strasbourg Observatory and 
the Universities of  Groningen, Heidelberg and Sydney. 
The RAVE web site is at \url{http://www.rave-survey.org}\\

\noindent \hrulefill

\noindent 
$^{1}$ Institut Utinam, CNRS UMR 6213, Universit\'e de Franche-Comt\'e, OSU THETA Franche-Comt\'e-Bourgogne, Observatoire de Be\-san\c{c}on, BP 1615, 25010 Besan\c{c}on Cedex, France.\\
$^{2}$ Centro de Investigaciones de Astronom\'ia, AP 264, M\'erida 5101-A, Venezuela.\\
$^{3}$ Dept. d'Astronomia i Meteorologia, Institut de Ci\`{e}ncies del Cosmos, Universitat de Barcelona (IEEC-UB), Mart\'{i} Franqu\`{e}s 1, E08028 Barcelona, Spain.\\
$^{4}$ Instituto de Astronom\'ia, Universidad Nacional Aut\'onoma de M\'exico, Apdo. Postal 70264, M\'exico D.F., 04510, Mexico.\\

\end{document}